# Energy spectra and the expectation values of diatomic molecules confined by the shifted Deng-Fan potential

Oluwadare[+1], O. J and Oyewumi[+2], K. J

[+1]Department of Physics, Federal University Oye-Ekiti, P. M. B. 373, Ekiti State, Nigeria
[+2]Theoretical Physics Section, Department of Physics, University of Ilorin, Ilorin, Nigeria.

**Abstract**

The approximate bound state solutions of the Schrodinger equation with shifted Deng-Fan potential was obtained via proper quantization rule. The energy spectra for the homogenous diatomic molecules ($H_2$, $I_2$); the heterogeneous diatomic molecules (CO, HCl, LiH); the neutral transition metal hydrides (ScH, TiH, VH, CrH); the transition-metal lithide (CuLi); the transition-metal carbides (TiC, NiC); the transition- metal nitrite (ScN) and the transition-metal fluoride (ScF) were calculated. By applying Hellmann-Feynman theorem, the expression for the expectation values of the square of inverse of position $r^{-2}$, potential energy V, kinetic energy T and the square of momentum $p^2$ were derived and the numerical values for diatomic molecules were presented. The result is reliable and very consistent with the available ones in the literature.

1. ## Introduction

The study on molecular dynamics and spectroscopy has been the subject of interest in recent time by some researchers in the field of chemistry and molecular physics. This is because it provides explanations about the dynamics and physical properties of some classes of molecules and that the potential energy involved are used to understand the bonding between two atoms [1-2]. The solutions of Schrödinger equation with different potential models of interest have been employed to proffer insight, explanations and predictions to the behaviour of some class of molecules [3-16]. One of these potentials is the shifted Deng-Fan potential model [15-16] which has been extensively used to explain bound and scattering states in quantum mechanics.

In literature, this shifted Deng-Fan potential has been used to obtain the approximate bound state solutions of the radial Schrödinger equation within the framework of parametric Nikiforov-Uvarov method [15]. The ro-vibrational energy levels for $H_2$, LiH, CO and HCl diatomic

**E-mail**: [+1]oluwatimilehin.oluwadare@fuoye.edu.ng; [+2]kjoyewumi66@unilorin.edu.ng

molecules were presented. Oyewumi et al. 2014 also studied the thermodynamics properties and the approximate solutions of the Schrödinger equation with this potential using asymptotic iteration method and applied their results to obtaining thermodynamic properties of some diatomic molecules [16].

In this present work, we are motivated to apply Hellmann – Feynman theorem to energy spectra and obtain the expectation values of some quantum mechanical observables ($r^{-2}$, V, T and $p^2$) for some classes of diatomic molecules (HCl, LiH, $H_2$, ScH, TiH, VH, CrH, CuLi, TiC, NiC, ScN, ScF & $I_2$) by using the shifted Deng-fan potential model as a tool for predictions.

The work is organized as follows; in Section 2, the proper quantization rules is briefly reviewed. The energy levels for the shifted Deng-Fan potential and the corresponding expectation values for some quantum mechanical observables are obtained and calculated numerically in Section 3. Finally, some concluding remarks are given in Section 4.

## 2. Overview of the Proper Quantization Rule

This method find its applications in various field of Physics including Mathematical, Nuclear, Atom, Chemical and Molecular Physics. The proper quantization rule is a simple, powerful and most suitable tools for obtaining bound state energy spectra of physically solvable potentials. Several authors, in different dimensions and contexts, have successfully applied this method to bound states problems [17-27].

The well-known one dimensional Schrödinger equation may be written as

$$\psi''(x) + k(x)^2 \psi(x) = 0, \quad k(x) = \sqrt{\frac{2\mu}{\hbar^2}[E - V(x)]}, \tag{2.1}$$

where the prime stand for the derivative with respect to variable $x$, $\mu$ denotes the reduced mass of the two non-relativistic particles, $k(x)$ is the exact wave number and $V(x)$ is a piecewise continuous real potential function of $x$. This equation can be written in form of non-linear Riccati equation

$$\phi'(x) + \phi(x)^2 + k(x) = 0, \tag{2.2}$$

where $\phi(x) = \psi'(x)/\psi(x)$ is the logarithmic derivative of the wave function $\psi(x)$.

According to Yang [19] in his explanation on monopoles: "For the Sturm-Liouville problem, the fundamental trick is the definition of a phase angle which is monotonic with respect to the energy". For the Schrödinger equation, the phase angle is the logarithmic derivative $\phi(x)$. From eq. (2.2), $\phi(x)$ decreases monotonically with respect to $x$ between two turning points, where $E \geq V(x)$. Specifically, as $x$ increases across a node of wave function $\psi(x)$, $\phi(x)$ decreases to $-\infty$, then jump to $+\infty$, and later decreases again.

Ma and Xu [17, 18] carefully studied one dimensional Schrödinger equation and generalized this exact quantization rule to a three dimensional radial Schrödinger equation with the spherically symmetric potential by making simple mappings $x \to r$ and $V(x) \to V_{eff}(r)$ as:

$$\int_{r_A}^{r_B} k(r)dr = N\pi + \int_{r_A}^{r_B} \left[\frac{dk(r)}{dr}\right]\left[\frac{d\phi(r)}{dr}\right]^{-1} \phi(r)dr, \quad k(r) = \sqrt{\frac{2\mu}{\hbar^2}[E - V_{eff}(r)]}, \quad (2.3)$$

where $r_A$ and $r_B$ are two turning points determined by $E = V_{eff}(r)$. The $N = n + 1$ is the number of nodes of $\phi(x)$ in the region $E \geq V_{eff}(r)$, and is larger by one than the number $n$ of nodes of the wave function $\psi(x)$. The first term $N\pi$ is the contribution from the nodes of the logarithmic derivative of wave function and the second is called the quantum correction.

Ma and Xu [17, 18] found that, for all well-known exactly solvable quantum systems, this quantum correction is independent of the number of nodes of wave function. This implies that quantum correction is the same for all bound states. Consequently, it is enough to consider the ground state in calculating the quantum correction

$$Q_0 = \left[\frac{dk_0'(r)}{dr}\right]\left[\frac{d\phi_0(r)}{dr}\right]^{-1} \phi_0(r)dr. \quad (2.4)$$

The calculations of two integrals in (2.3) and (2.4), particularly, the calculation of quantum correction term are very difficult and tedious for some physical potentials. To overcome this, Serrano *et al* [26-27] have proposed proper quantization rule as follows:

$$\int_{r_{0A}}^{r_{0B}} k_0(r)dr = \pi + \int_{r_{0A}}^{r_{0B}} \left[\frac{dk_0(r)}{dr}\right]\left[\frac{d\phi_0(r)}{dr}\right]^{-1} \phi_0(r)dr, \quad k_0(r) = \sqrt{\frac{2\mu}{\hbar^2}[E_0 - V_{eff}(r)]}, \quad (2.5)$$

where they have taking $N = 1$, i.e., $n = 0$ in eq. (2.3). Consequently, the complicated quantum correction turns to

$$\int_{r_{0A}}^{r_{0B}} \left[\frac{dk_0(r)}{dr}\right]\left[\frac{d\phi_0(r)}{dr}\right]^{-1} \phi_0(r)dr = \int_{r_{0A}}^{r_{0B}} k_0(r)dr - \pi. \quad (2.6)$$

Finally, by inserting Eq. (2.6) into eq. (2.3), we obtain the following integral

$$\int_{r_A}^{r_B} k(r)dr - \int_{r_{0A}}^{r_{0B}} k_0(r)dr = (N-1)\pi = n\pi. \quad (2.7)$$

Eq. (2.7) is the well known as the proper quantization rule. The two integrals involved in proper quantization rule have the same mathematical form. Accordingly, when applying it to calculate the energy levels, we can calculate its first integral with respect to $k(r)$, and then replace energy levels $E_n$ in the result by the ground state energy $E_0$ to obtain the second integral. This will greatly simplify the complicated integral calculations occurred previously [17-27].

### 3.1. Application of Proper Quantization Rule to radial Schrödinger equation with spherically symmetric shifted Deng-Fan potential

The shifted Deng-Fan exponential-type potential that is under investigation is defined [16] as

$$V(r) = D\left[1 - \frac{b}{e^{ar}-1}\right]^2 - D, \quad b = e^{ar_e} - 1, \quad r \forall [0, \infty) \tag{3.1}$$

where $D$ is the dissociation energy, $r_e$ is the equilibrium inter-nuclear distance, $a$ is the radius or range of the potential and $b$ is the position of minimum $r_e$.

The radial Schrödinger equation with this potential can be expressed as:

$$\frac{d^2 R(r)}{dr^2} + \frac{2\mu}{\hbar^2}\left[E - V_{eff}\right]R(r) = 0, \tag{3.2}$$

with the effective potential

$$V_{eff} = V(r) + \frac{l(l+1)\hbar^2}{2\mu r^2}, \tag{3.3}$$

where $\mu$ and $V(r)$ stand for the reduced mass and interaction potential respectively.

Putting Eqs. (3.1) and (3.3) into Eq. (3.2), we obtain

$$\frac{d^2 R(r)}{dr^2} + \frac{2\mu}{\hbar^2}\left[E - D\left(1 - \frac{b}{e^{ar}-1}\right)^2 + D - \frac{l(l+1)\hbar^2}{2\mu r^2}\right]R(r) = 0. \tag{3.4}$$

Using an approximation of the type [16, 28-31]

$$\frac{1}{r^2} \approx a^2\left[C_0 + \frac{e^{ar}}{(e^{ar}-1)^2}\right] \tag{3.5}$$

to deal with the centrifugal term, where the dimensionless parameter $C_0 = \frac{1}{12}$ by the above series expansion. By introducing a variable $\xi = (e^{ar} - 1)^{-1}$ and apply the above approximation in Eq. (3.4), we have

$$\frac{d^2 R(\xi)}{d\xi^2} + \frac{2\mu}{\hbar^2}\left[E - V_{eff}(\xi)\right]R(\xi) = 0, \tag{3.6}$$

where

$$V_{eff}(\xi) = P + Q\xi + R\xi^2, \quad P = \frac{l(l+1)a^2\hbar^2 C_0}{2\mu}, \quad Q = \frac{l(l+1)a^2\hbar^2}{2\mu} - 2bD, \quad R = \frac{l(l+1)a^2\hbar^2}{2\mu} + Db^2. \tag{3.7}$$

The two turning points $\xi_A$ and $\xi_B$ are obtained by solving for $V_{eff}(\xi) - E_{nl} = 0$:

$$\xi_A = -\frac{Q}{2R} - \frac{1}{2R}\sqrt{Q^2 - 4R(P - E)}, \quad \xi_B = -\frac{Q}{2R} + \frac{1}{2R}\sqrt{Q^2 - 4R(P - E)}, \tag{3.8}$$

with the following properties

$$\xi_A + \xi_B = -\frac{Q}{R}, \quad \xi_A \xi_B = \frac{P-E}{R}. \tag{3.8}$$

The momentum $k(\xi)$ between the two turning points can be evaluated as

$$k(y) = \sqrt{\frac{2\mu R}{\hbar^2} [(-1)(\xi - \xi_A)(\xi - \xi_B)]^2}. \tag{3.9}$$

The non-linear Riccati equation for the ground state (2.2) can be written as:

$$-a\xi(1+\xi)\phi_0'(\xi) = -\frac{2\mu}{\hbar^2}[E_0 - V_{eff}(\xi)] - [\phi_0(\xi)]^2. \tag{3.10}$$

By considering the monotonic property, the Logarithmic derivative of $\phi_0(\xi)$ for the ground state has one node and no pole, so it has to take a linear form in $y$. Therefore, we assume a trial solution of the form $\phi_0(\xi) = A + B\xi$ and substituting it into Eq. (3.10) we obtain the following equation

$$A^2 + (-aB + 2AB)\xi + (-aB + B^2)\xi^2 = -\frac{2\mu}{\hbar^2}(E_0 - P) + \frac{2\mu Q}{\hbar^2}\xi + \frac{2\mu R}{\hbar^2}\xi^2. \tag{3.11}$$

Comparing the LHS and RHS of equation (3.11) and solving for $B$, we obtain the values of $A$ and $B$ as:

$$A = \sqrt{-\frac{2\mu}{\hbar^2}(E_0 - P)}, \quad B = \left[\frac{a}{2} \pm \sqrt{\frac{a^2}{4} + \frac{2\mu R}{\hbar^2}}\right], \tag{3.12}$$

while the problem can only be physically solvable for $B = \left[\frac{a}{2} + \sqrt{\frac{a^2}{4} + \frac{2\mu R}{\hbar^2}}\right]$. Thus, the integral of the momentum $k(r)$ in terms of $\xi$ is

$$\int_{r_A}^{r_B} k(r) dr = -\int_{\xi_A}^{\xi_B} \left\{\frac{1}{a}\sqrt{\frac{2\mu R}{\hbar^2}}\left[\frac{\sqrt{(-1)(\xi - \xi_A)(\xi - \xi_B)}}{\xi(1+\xi)}\right]\right\} d\xi. \tag{3.13}$$

In order to obtain the energy levels of all the bound states, we consider the following useful integrals [26]

$$\int_{\xi_A}^{\xi_B} \left\{\left[\frac{\sqrt{(-1)(\xi - \xi_A)(\xi - \xi_B)}}{\xi(1+\xi)}\right]\right\} d\xi = \pi\left[\sqrt{(\xi_A + 1)(\xi_B + 1)} - 1 - \sqrt{\xi_A \xi_B}\right]. \tag{3.14}$$

Consequently, the integral of the momentum $k(r)$ and $k_0(r)$ are thus evaluated respectively as:

$$\int_{r_A}^{r_B} k(r) dr = -\frac{\pi}{a}\sqrt{\frac{2\mu R}{\hbar^2}}\left[\left(\frac{R-Q+P-E_0}{R}\right)^{1/2} - 1 - \left(\frac{P-E_0}{R}\right)^{1/2}\right] \tag{3.15}$$

$$\int_{r_{0A}}^{r_{0B}} k_0(r) dr = \frac{\pi}{a}\left[B + \sqrt{\frac{2\mu R}{\hbar^2}}\right] \tag{3.16}$$

The proper quantization rule in Eq. (2.9) requires that $E_0$ in eq. (3.15) be replaced with $E_n$, therefore, we obtain the explicit energy levels of all the bound states as:

$$E_{n,l} = \frac{l(l+1)a^2\hbar^2 C_0}{2\mu} - \frac{a^2\hbar^2}{2\mu}\left[\frac{\frac{2\mu b}{a^2\hbar^2}(2+b)D-(n+\rho/a)^2}{2(n+\rho/a)}\right]^2, \tag{3.17}$$

where $\rho = \frac{a}{2}\left[1+\sqrt{(1+2l)^2 + \frac{8\mu D b^2}{\hbar^2}}\right]$ and $n$ is the number of nodes of the wave function.

### 3.2. Application of Hellmann – Feynman theorem to the shifted Deng-Fan potential

Hellmann – Feynman Theorem (HFT) is one of the useful means of obtaining expectation values of some quantum mechanical observables for any arbitrary values of $n$ and $l$ quantum numbers. Assuming that the Hamiltonian $\hat{H}$ for a particular quantum mechanical system is a function of some parameter q, let E (q) and $\Psi(q)$ be the eigenvalues and the eigenfunctions of the Hamiltonian $\hat{H}(q)$. Then, the Hellmann – Feynman Theorem (HFT) states that

$$\frac{\partial E_{nl}(q)}{\partial q} = \left\langle \Psi_{nl}(q) \left| \frac{\partial \hat{H}(q)}{\partial q} \right| \Psi_{nl}(q) \right\rangle, \tag{3.18}$$

provided that the associated normalized eigenfunctions $\Psi_{nl}(q)$ is continuous with respect to the parameter $q$ [32-41].

The Shifted Deng – Fan molecular potential given in (3.1) may be written as

$$V_{DF}(r) = D\left[\frac{e^{ar}-e^{ar_e}}{e^{ar}-1}\right]^2 - D \tag{3.19}$$

The effective Hamiltonian of the Shifted Deng- Fan potential radial wave function is

$$\hat{H} = -\frac{\hbar^2}{2\mu}\frac{d^2}{dr^2} + \frac{\hbar^2}{2\mu}\frac{l(l+1)}{r^2} + D\left[\frac{e^{ar}-e^{ar_e}}{e^{ar}-1}\right]^2 - D. \tag{3.20}$$

With $= l$, $q = D$, and $q = \mu$, we apply Eq. (3.18) and then obtain the following expectation values of $r^{-2}$, $V$, $P^2$ and $T$ respectively as:

$$\langle r^{-2} \rangle_{nl} = a^2\left(C_0 - \frac{2\omega\chi}{1+2l}\right), \tag{3.21}$$

where

$$\omega = \left[\frac{\frac{2\mu}{\hbar^2 a^2}b(2+b)D-(n+\eta)^2}{2(n+\eta)}\right], \tag{3.22}$$

$$\chi = -\left[\frac{\frac{4\mu}{\hbar^2 a^2}b(2+b)\lambda D+2\lambda(n+\eta)^2}{4(n+\eta)^2}\right], \tag{3.23}$$

$$\lambda = (1+2l)\left[(1+2l)^2 + \frac{8\mu D b^2}{\hbar^2 a^2}\right]^{-\frac{1}{2}}, \tag{3.24}$$

$$\langle \hat{V} \rangle_{nl} = -\frac{\hbar^2 a^2}{\mu}D\omega\Omega, \tag{3.25}$$

with the following dimensional parameters

$$\Omega = \left\{ \frac{\frac{4\mu}{\hbar^2 a^2} b(2+b)[(n+\eta)-D\zeta]-2\zeta(n+\eta)^2}{4(n+\eta)^2} \right\}, \zeta = \frac{2\mu b^2}{\hbar^2 a^2}\left[(1+2l)^2 + \frac{8\mu D b^2}{\hbar^2 a^2}\right]^{-\frac{1}{2}}. \quad (3.26)$$

$$\langle \widehat{P^2} \rangle_{nl} = \hbar^2 l(l+1) a^2 C_0 - \omega \hbar^2 a^2 (\omega - 2\mu\gamma), \quad (3.27)$$

$$\langle \widehat{T} \rangle_{nl} = \frac{\hbar^2 l(l+1) a^2 C_0}{2\mu} + \frac{\omega \hbar^2 a^2}{2\mu}(\omega - 2\mu\gamma), \quad (3.28)$$

where

$$\gamma = \left\{ \frac{\frac{4bD}{\hbar^2 a^2}(2+b)[(n+\eta)-\beta\mu]-2\beta(n+\eta)^2}{4(n+\eta)^2} \right\}; \beta = \frac{2D b^2}{\hbar^2 a^2}\left[(1+2l)^2 + \frac{8\mu D b^2}{\hbar^2 a^2}\right]^{-\frac{1}{2}}. \quad (3.29)$$

## 4. Numerical Results

**Table 1**: Model parameters for some selected diatomic molecules in this study

| Molecules | $D_e(eV)$ | $r_e(Å)$ | $a(Å^{-1})$ | $\mu$ (a. m. u) |
|---|---|---|---|---|
| HCl | 4.619030905 | 1.2746 | 1.8677 | 0.9801045 |
| LiH | 2.5152672118 | 1.5956 | 1.1280 | 0.8801221 |
| $H_2$ | 4.7446 | 0.7416 | 1.9426 | 0.50391 |
| ScH | 2.25 | 1.776 | 1.41113 | 0.986040 |
| ScN | 4.56 | 1.768 | 1.50680 | 10.682771 |
| TiH | 2.05 | 1.781 | 1.32408 | 0.987371 |
| VH | 2.33 | 1.719 | 1.44370 | 0.988005 |
| CrH | 2.13 | 1.694 | 1.52179 | 0.988976 |
| NiC | 2.76 | 1.621 | 2.25297 | 9.974265 |
| CuLi | 1.74 | 2.310 | 1.00818 | 6.259494 |
| TiC | 2.66 | 1.790 | 1.52550 | 9.606079 |
| ScF | 5.85 | 1.794 | 1.46102 | 13.358942 |
| CO | 11.2256 | 1.1283 | 2.2994 | 6.8606719 |
| $I_2$ | 1.5556 | 2.662 | 1.8643 | 63.45223502 |

The spectroscopic parameters are taken from [12, 16], also, the following conversion 1amu=931.494028$MeV/c^2$ [12, 16] and $\hbar c = 1973.29\ eV$Å [12, 16] are used throughout the computations.

Table 2: The energy spectra $E_{n,l}$(eV) and expectation values of $\langle r^{-2} \rangle_{nl}$, $\langle V \rangle_{nl}$, $\langle T \rangle_{nl}$ and $\langle p^2 \rangle_{nl}$ corresponding to the shifted Deng – Fan molecular potential with various $n$ and $l$ quantum numbers for ScH diatomic molecule.

| ScH | | | | | | |
|---|---|---|---|---|---|---|
| $n$ | $l$ | $\langle r^{-2} \rangle_{nl}$ (Å$^{-2}$) | $\langle V \rangle_{nl}$(eV) | $\langle T \rangle_{nl}$(eV) | $E_{n,l}$(eV) | $\langle p^2 \rangle_{nl}$((eV/c)$^2$) |
| 0 | 0 | 0.3536004243 | -2.196945170 | 0.05179586338 | -2.145149306 | 0.000000001057200040 |
| 0 | 1 | 0.3535598565 | -2.196934652 | 0.05328432273 | -2.143650329 | 0.000000001087580830 |
| 0 | 2 | 0.3534787318 | -2.196913102 | 0.05626021162 | -2.140652890 | 0.000000001148321393 |
| 0 | 3 | 0.3533570716 | -2.196879492 | 0.06072147110 | -2.136158021 | 0.000000001239379701 |
| 0 | 4 | 0.3531949084 | -2.196832284 | 0.06666501448 | -2.130167269 | 0.000000001360692754 |
| 0 | 5 | 0.3529922852 | -2.196769425 | 0.07408672934 | -2.122682696 | 0.000000001512176612 |
| 1 | 0 | 0.3438378507 | -2.090868820 | 0.1471153578 | -1.943753462 | 0.000000003002756435 |
| 1 | 1 | 0.3437978902 | -2.090837865 | 0.1485419937 | -1.942295871 | 0.000000003031875354 |
| 1 | 2 | 0.3437179799 | -2.090775452 | 0.1513942549 | -1.939381198 | 0.000000003090092564 |
| 1 | 3 | 0.3435981409 | -2.090680578 | 0.1556701209 | -1.935010457 | 0.000000003177366826 |
| 1 | 4 | 0.3434384051 | -2.090551737 | 0.1613665628 | -1.929185174 | 0.000000003293636315 |
| 1 | 5 | 0.3432388149 | -2.090386923 | 0.1684795449 | -1.921907378 | 0.000000003438818661 |
| 2 | 0 | 0.3341944352 | -1.984890801 | 0.2318414882 | -1.753049313 | 0.000000004732092769 |
| 2 | 1 | 0.3341550715 | -1.984839907 | 0.2332073037 | -1.751632603 | 0.000000004759970288 |
| 2 | 2 | 0.3340763546 | -1.984737627 | 0.2359379428 | -1.748799684 | 0.000000004815705083 |
| 2 | 3 | 0.3339583052 | -1.984582981 | 0.2400314229 | -1.744551558 | 0.000000004899256684 |
| 2 | 4 | 0.3338009548 | -1.984374495 | 0.2454847714 | -1.738889724 | 0.000000005010564419 |
| 2 | 5 | 0.3336043449 | -1.984110210 | 0.2522940278 | -1.731816183 | 0.000000005149547451 |
| 3 | 0 | 0.3246679247 | -1.879073664 | 0.3061668623 | -1.572906802 | 0.000000006249140335 |
| 3 | 1 | 0.3246291475 | -1.879003315 | 0.3074728367 | -1.571530478 | 0.000000006275796444 |
| 3 | 2 | 0.3245516033 | -1.878862135 | 0.3100838122 | -1.568778323 | 0.000000006329088796 |
| 3 | 3 | 0.3244353127 | -1.878649166 | 0.3139978429 | -1.564651324 | 0.000000006408977673 |
| 3 | 4 | 0.3242803064 | -1.878362969 | 0.3192120115 | -1.559150958 | 0.000000006515403532 |
| 3 | 5 | 0.3240866254 | -1.878001628 | 0.3257224313 | -1.552279196 | 0.000000006648287042 |

Table 3: The energy spectra $E_{n,l}$(eV) and expectation values of $\langle r^{-2} \rangle_{nl}$, $\langle V \rangle_{nl}$, $\langle T \rangle_{nl}$ and $\langle p^2 \rangle_{nl}$ corresponding to the shifted Deng – Fan molecular potential with various $n$ and $l$ quantum numbers for ScN diatomic molecule.

| ScN | | | | | | |
|---|---|---|---|---|---|---|
| $n$ | $l$ | $\langle r^{-2} \rangle_{nl}$ (Å$^{-2}$) | $\langle V \rangle_{nl}$(eV) | $\langle T \rangle_{nl}$(eV) | $E_{n,l}$(eV) | $\langle p^2 \rangle_{nl}$((eV/c)$^2$) |
| 0 | 0 | 0.3709156088 | -4.535811235 | 0.02405983855 | -4.511751396 | 0.000000005320398822 |
| 0 | 1 | 0.3709139715 | -4.535811031 | 0.02420477621 | -4.511606254 | 0.000000005352449168 |
| 0 | 2 | 0.3709106970 | -4.535810621 | 0.02449464769 | -4.511315973 | 0.000000005416549012 |
| 0 | 3 | 0.3709057853 | -4.535810001 | 0.02492944530 | -4.510880556 | 0.000000005512696653 |
| 0 | 4 | 0.3708992364 | -4.535809166 | 0.02550915753 | -4.510300009 | 0.000000005640889542 |
| 0 | 5 | 0.3708910505 | -4.535808108 | 0.02623376898 | -4.509574339 | 0.000000005801124281 |
| 1 | 0 | 0.3688371872 | -4.487434353 | 0.07133635241 | -4.416098001 | 0.00000001577474614 |
| 1 | 1 | 0.3688355547 | -4.487433744 | 0.07148007108 | -4.415953673 | 0.00000001580652693 |
| 1 | 2 | 0.3688322897 | -4.487432522 | 0.07176750459 | -4.415665018 | 0.00000001587008766 |
| 1 | 3 | 0.3688273923 | -4.487430686 | 0.07219864529 | -4.415232041 | 0.00000001596542664 |
| 1 | 4 | 0.3688208625 | -4.487428228 | 0.07277348169 | -4.414654746 | 0.00000001609254133 |
| 1 | 5 | 0.3688127003 | -4.487425142 | 0.07349199847 | -4.413933143 | 0.00000001625142834 |
| 2 | 0 | 0.3667636575 | -4.439059414 | 0.1175157234 | -4.321543690 | 0.00000002598648013 |
| 2 | 1 | 0.3667620297 | -4.439058400 | 0.1176582269 | -4.321400173 | 0.00000002601799221 |
| 2 | 2 | 0.3667587742 | -4.439056372 | 0.1179432301 | -4.321113141 | 0.00000002608101552 |
| 2 | 3 | 0.3667538910 | -4.439053324 | 0.1183707254 | -4.320682598 | 0.00000002617554837 |
| 2 | 4 | 0.3667473801 | -4.439049251 | 0.1189407012 | -4.320108550 | 0.00000002630158824 |
| 2 | 5 | 0.3667392417 | -4.439044146 | 0.1196531425 | -4.319391003 | 0.00000002645913175 |
| 3 | 0 | 0.3646950014 | -4.390687699 | 0.1626018197 | -4.228085879 | 0.00000003595645615 |
| 3 | 1 | 0.3646933784 | -4.390686284 | 0.1627431118 | -4.227943172 | 0.00000003598770036 |
| 3 | 2 | 0.3646901324 | -4.390683451 | 0.1630256923 | -4.227657759 | 0.00000003605018793 |
| 3 | 3 | 0.3646852633 | -4.390679198 | 0.1634495535 | -4.227229644 | 0.00000003614391718 |
| 3 | 4 | 0.3646787714 | -4.390673517 | 0.1640146841 | -4.226658833 | 0.00000003626888560 |
| 3 | 5 | 0.3646706565 | -4.390666403 | 0.1647210688 | -4.225945334 | 0.00000003642508982 |

Table 4: The energy spectra $E_{n,l}$(eV) and expectation values of $\langle r^{-2}\rangle_{nl}$, $\langle V\rangle_{nl}$, $\langle T\rangle_{nl}$ and $\langle p^2\rangle_{nl}$ corresponding to the shifted Deng – Fan molecular potential with various $n$ and $l$ quantum numbers for TiH diatomic molecule.

| TiH | | | | | | |
|---|---|---|---|---|---|---|
| $n$ | $l$ | $\langle r^{-2}\rangle_{nl}$ (Å$^{-2}$) | $\langle V\rangle_{nl}$(eV) | $\langle T\rangle_{nl}$(eV) | $E_{n,l}$(eV) | $\langle p^2\rangle_{nl}$((eV/c)$^2$) |
| 0 | 0 | 0.3431643269 | -2.001831705 | 0.04702684061 | -1.954804865 | 0.0000000009611556677 |
| 0 | 1 | 0.3431141376 | -2.001820549 | 0.04846843726 | -1.953352112 | 0.0000000009906196668 |
| 0 | 2 | 0.3430137767 | -2.001797602 | 0.05135035864 | -1.950447243 | 0.0000000010495216690 |
| 0 | 3 | 0.3428632799 | -2.001761596 | 0.05567006205 | -1.946091534 | 0.0000000011378097050 |
| 0 | 4 | 0.3426627006 | -2.001710630 | 0.06142373628 | -1.940286894 | 0.0000000012554058810 |
| 0 | 5 | 0.3424121100 | -2.001642174 | 0.06860630493 | -1.933035869 | 0.0000000014022064410 |
| 1 | 0 | 0.3328234874 | -1.905530701 | 0.1334798045 | -1.772050896 | 0.000000002728120133 |
| 1 | 1 | 0.3327741627 | -1.905497972 | 0.1348560499 | -1.770641922 | 0.000000002756248453 |
| 1 | 2 | 0.3326755309 | -1.905431895 | 0.1376072960 | -1.767824599 | 0.000000002812479656 |
| 1 | 3 | 0.3325276267 | -1.905331235 | 0.1417310549 | -1.763600180 | 0.000000002896762890 |
| 1 | 4 | 0.3323305027 | -1.905194141 | 0.1472235972 | -1.757970544 | 0.000000003009021934 |
| 1 | 5 | 0.3320842285 | -1.905018147 | 0.1540799557 | -1.750938191 | 0.000000003149155265 |
| 2 | 0 | 0.3226295849 | -1.809334477 | 0.2102588924 | -1.599075584 | 0.000000004297365580 |
| 2 | 1 | 0.3225811076 | -1.809280786 | 0.2115710203 | -1.597709766 | 0.000000004324183438 |
| 2 | 2 | 0.3224841700 | -1.809172803 | 0.2141940581 | -1.594978745 | 0.000000004377794262 |
| 2 | 3 | 0.3223388066 | -1.809009323 | 0.2181255712 | -1.590883752 | 0.000000004458148290 |
| 2 | 4 | 0.3221450686 | -1.808788543 | 0.2233619102 | -1.585426633 | 0.000000004565170936 |
| 2 | 5 | 0.3219030244 | -1.808508061 | 0.2298982144 | -1.578609846 | 0.000000004698762853 |
| 3 | 0 | 0.3125794176 | -1.713309199 | 0.2775686861 | -1.435740513 | 0.000000005673073343 |
| 3 | 1 | 0.3125317707 | -1.713235139 | 0.2788178968 | -1.434417242 | 0.000000005698605270 |
| 3 | 2 | 0.3124364938 | -1.713086432 | 0.2813151262 | -1.431771306 | 0.000000005749644765 |
| 3 | 3 | 0.3122936204 | -1.712861906 | 0.2850579917 | -1.427803914 | 0.000000005826143130 |
| 3 | 4 | 0.3121032010 | -1.712559803 | 0.2900429221 | -1.422516881 | 0.000000005928027374 |
| 3 | 5 | 0.3118653025 | -1.712177783 | 0.2962651605 | -1.415912623 | 0.000000006055200273 |

Table 5: The energy spectra $E_{n,l}$(eV) and expectation values of $\langle r^{-2}\rangle_{nl}$, $\langle V\rangle_{nl}$, $\langle T\rangle_{nl}$ and $\langle p^2\rangle_{nl}$ corresponding to the shifted Deng – Fan molecular potential with various $n$ and $l$ quantum numbers for VH diatomic molecule.

| VH | | | | | | |
|---|---|---|---|---|---|---|
| $n$ | $l$ | $\langle r^{-2}\rangle_{nl}$ (Å$^{-2}$) | $\langle V\rangle_{nl}$(eV) | $\langle T\rangle_{nl}$(eV) | $E_{n,l}$(eV) | $\langle p^2\rangle_{nl}$((eV/c)$^2$) |
| 0 | 0 | 0.3758143282 | -2.274697476 | 0.05398117627 | -2.220716299 | 0.000000001103999934 |
| 0 | 1 | 0.3757687921 | -2.274686066 | 0.05555973951 | -2.219126327 | 0.000000001136284034 |
| 0 | 2 | 0.3756777327 | -2.274662672 | 0.05871571245 | -2.215946960 | 0.000000001200828643 |
| 0 | 3 | 0.3755411758 | -2.274626142 | 0.06344678877 | -2.211179353 | 0.000000001297586593 |
| 0 | 4 | 0.3753591600 | -2.274574752 | 0.06974951106 | -2.204825241 | 0.000000001426487174 |
| 0 | 5 | 0.3751317369 | -2.274506204 | 0.07761927325 | -2.196886931 | 0.000000001587436185 |
| 1 | 0 | 0.3652091010 | -2.164128597 | 0.1532439972 | -2.010884599 | 0.000000003134080702 |
| 1 | 1 | 0.3651642675 | -2.164095040 | 0.1547555436 | -2.009339496 | 0.000000003164994202 |
| 1 | 2 | 0.3650746132 | -2.164027363 | 0.1577775051 | -2.006249858 | 0.000000003226798066 |
| 1 | 3 | 0.3649401634 | -2.163924444 | 0.1623076197 | -2.001616824 | 0.000000003319446033 |
| 1 | 4 | 0.3647609561 | -2.163784596 | 0.1683424967 | -1.995442099 | 0.000000003442868756 |
| 1 | 5 | 0.3645370418 | -2.163605575 | 0.1758776186 | -1.987727957 | 0.000000003596973847 |
| 2 | 0 | 0.3547375366 | -2.053666433 | 0.2413781874 | -1.812288246 | 0.000000004936563472 |
| 2 | 1 | 0.3546933933 | -2.053611286 | 0.2428238396 | -1.810787446 | 0.000000004966129332 |
| 2 | 2 | 0.3546051189 | -2.053500441 | 0.2457140344 | -1.807786407 | 0.000000005025238359 |
| 2 | 3 | 0.3544727385 | -2.053332802 | 0.2500465533 | -1.803286248 | 0.000000005113845183 |
| 2 | 4 | 0.3542962894 | -2.053106721 | 0.2558180706 | -1.797288651 | 0.000000005231881788 |
| 2 | 5 | 0.3540758213 | -2.052820008 | 0.2630241559 | -1.789795852 | 0.000000005379257561 |
| 3 | 0 | 0.3443970295 | -1.943378770 | 0.3185926423 | -1.624786127 | 0.000000006515720484 |
| 3 | 1 | 0.3443535639 | -1.943302572 | 0.3199734956 | -1.623329077 | 0.000000006543961105 |
| 3 | 2 | 0.3442666450 | -1.943149641 | 0.3227341138 | -1.620415527 | 0.000000006600420089 |
| 3 | 3 | 0.3441362972 | -1.942918904 | 0.3268723211 | -1.616046582 | 0.000000006685052935 |
| 3 | 4 | 0.3439625572 | -1.942608751 | 0.3323848556 | -1.610223895 | 0.000000006797792932 |
| 3 | 5 | 0.3437454738 | -1.942217042 | 0.3392673715 | -1.602949671 | 0.000000006938551204 |

Table 6: The energy spectra $E_{n,l}$(eV) and expectation values of $\langle r^{-2} \rangle_{nl}$, $\langle V \rangle_{nl}$, $\langle T \rangle_{nl}$ and $\langle p^2 \rangle_{nl}$ corresponding to the shifted Deng – Fan molecular potential with various $n$ and $l$ quantum numbers for CrH diatomic molecule.

| CrH | | | | | | |
|---|---|---|---|---|---|---|
| $n$ | $l$ | $\langle r^{-2} \rangle_{nl}$ (Å$^{-2}$) | $\langle V \rangle_{nl}$(eV) | $\langle T \rangle_{nl}$(eV) | $E_{n,l}$(eV) | $\langle p^2 \rangle_{nl}$((eV/c)$^2$) |
| 0 | 0 | 0.3931558610 | -2.074753734 | 0.05380549333 | -2.02094824 | 0.000000001101488409 |
| 0 | 1 | 0.3931075278 | -2.074741472 | 0.05545493765 | -2.019286535 | 0.000000001135255292 |
| 0 | 2 | 0.3930108748 | -2.074716340 | 0.05875260307 | -2.015963736 | 0.000000001202764016 |
| 0 | 3 | 0.3928659291 | -2.074677115 | 0.06369604398 | -2.010981071 | 0.000000001303964517 |
| 0 | 4 | 0.3926727310 | -2.074621970 | 0.07028159409 | -2.004340376 | 0.000000001438781739 |
| 0 | 5 | 0.3924313344 | -2.074548468 | 0.07850436900 | -1.996044099 | 0.000000001607115689 |
| 1 | 0 | 0.3817450347 | -1.964299844 | 0.1520153064 | -1.812284538 | 0.000000003112007484 |
| 1 | 1 | 0.3816974429 | -1.964263772 | 0.1535927097 | -1.810671062 | 0.000000003144299567 |
| 1 | 2 | 0.3816022726 | -1.964191032 | 0.1567463165 | -1.807444715 | 0.000000003208859172 |
| 1 | 3 | 0.3814595502 | -1.964080431 | 0.1614737281 | -1.802606703 | 0.000000003305637190 |
| 1 | 4 | 0.3812693155 | -1.963930182 | 0.1677713482 | -1.796158834 | 0.000000003434560004 |
| 1 | 5 | 0.3810316213 | -1.963737907 | 0.1756343861 | -1.788103520 | 0.000000003595529536 |
| 2 | 0 | 0.3704761060 | -1.853959982 | 0.2381706979 | -1.615789284 | 0.000000004875752397 |
| 2 | 1 | 0.3704292424 | -1.853900691 | 0.2396772509 | -1.614223440 | 0.000000004906594056 |
| 2 | 2 | 0.3703355284 | -1.853781524 | 0.2426891801 | -1.611092344 | 0.000000004968253281 |
| 2 | 3 | 0.3701949899 | -1.853601318 | 0.2472041323 | -1.606397186 | 0.000000005060681902 |
| 2 | 4 | 0.3700076660 | -1.853358327 | 0.2532185803 | -1.600139747 | 0.000000005183807708 |
| 2 | 5 | 0.3697736086 | -1.853050225 | 0.2607278246 | -1.592322401 | 0.000000005337534494 |
| 3 | 0 | 0.3593463287 | -1.743806604 | 0.3124949130 | -1.431311691 | 0.000000006397293345 |
| 3 | 1 | 0.3593001807 | -1.743724666 | 0.3139317776 | -1.429792889 | 0.000000006426708366 |
| 3 | 2 | 0.3592078973 | -1.743560221 | 0.3168043524 | -1.426755869 | 0.000000006485514775 |
| 3 | 3 | 0.3590695042 | -1.743312130 | 0.3211103292 | -1.422201801 | 0.000000006573665320 |
| 3 | 4 | 0.3588850397 | -1.742978689 | 0.3268462479 | -1.416132441 | 0.000000006691089167 |
| 3 | 5 | 0.3586545550 | -1.742557623 | 0.3340074987 | -1.408550125 | 0.000000006837691943 |

Table 7: The energy spectra $E_{n,l}$(eV) and expectation values of $\langle r^{-2} \rangle_{nl}$, $\langle V \rangle_{nl}$, $\langle T \rangle_{nl}$ and $\langle p^2 \rangle_{nl}$ corresponding to the shifted Deng – Fan molecular potential with various $n$ and $l$ quantum numbers for NiC diatomic molecule.

| NiC | | | | | | |
|---|---|---|---|---|---|---|
| $n$ | $l$ | $\langle r^{-2} \rangle_{nl}$ (Å$^{-2}$) | $\langle V \rangle_{nl}$(eV) | $\langle T \rangle_{nl}$(eV) | $E_{n,l}$(eV) | $\langle p^2 \rangle_{nl}$((eV/c)$^2$) |
| 0 | 0 | 0.5603148568 | -2.732187594 | 0.02753195653 | -2.704655638 | 0.000000005684411885 |
| 0 | 1 | 0.5603133197 | -2.732187293 | 0.02776648492 | -2.704420808 | 0.000000005732834014 |
| 0 | 2 | 0.5603102455 | -2.732186689 | 0.02823553783 | -2.703951152 | 0.000000005829677476 |
| 0 | 3 | 0.5603056342 | -2.732185779 | 0.02893910754 | -2.703246671 | 0.000000005974940673 |
| 0 | 4 | 0.5602994859 | -2.732184556 | 0.02987718246 | -2.702307373 | 0.000000006168621213 |
| 0 | 5 | 0.5602918005 | -2.732183012 | 0.03104974713 | -2.701133265 | 0.000000006410715905 |
| 1 | 0 | 0.5574484820 | -2.676563706 | 0.08085776860 | -2.595705938 | 0.000000016694376960 |
| 1 | 1 | 0.5574469481 | -2.676562806 | 0.08109049600 | -2.595472310 | 0.000000016742427250 |
| 1 | 2 | 0.5574438803 | -2.676561002 | 0.08155594693 | -2.595005055 | 0.000000016838527020 |
| 1 | 3 | 0.5574392786 | -2.676558293 | 0.08225411371 | -2.594304179 | 0.00000001698267470 |
| 1 | 4 | 0.5574331430 | -2.676554671 | 0.08318498475 | -2.593369686 | 0.00000001717486788 |
| 1 | 5 | 0.5574254736 | -2.676550129 | 0.08434854465 | -2.592201584 | 0.00000001741510340 |
| 2 | 0 | 0.5545867053 | -2.620942582 | 0.1318901057 | -2.489052477 | 0.00000002723081753 |
| 2 | 1 | 0.5545851746 | -2.620941084 | 0.1321210360 | -2.488820048 | 0.00000002727849676 |
| 2 | 2 | 0.5545821131 | -2.620938085 | 0.1325828926 | -2.488355193 | 0.00000002737385444 |
| 2 | 3 | 0.5545775210 | -2.620933582 | 0.1332756680 | -2.487657914 | 0.00000002751688898 |
| 2 | 4 | 0.5545713981 | -2.620927569 | 0.1341993506 | -2.486728219 | 0.00000002770759799 |
| 2 | 5 | 0.5545637446 | -2.620920039 | 0.1353539250 | -2.485566114 | 0.00000002794597830 |
| 3 | 0 | 0.5517295147 | -2.565326054 | 0.1806344807 | -2.384691573 | 0.00000003729487179 |
| 3 | 1 | 0.5517279871 | -2.565323960 | 0.1808636176 | -2.384460342 | 0.00000003734218077 |
| 3 | 2 | 0.5517249320 | -2.565319769 | 0.1813218877 | -2.383997882 | 0.00000003743679794 |
| 3 | 3 | 0.5517203494 | -2.565313479 | 0.1820092832 | -2.383304196 | 0.00000003757872171 |
| 3 | 4 | 0.5517142392 | -2.565305083 | 0.1829257926 | -2.382379290 | 0.00000003776794972 |
| 3 | 5 | 0.5517066016 | -2.565294573 | 0.1840714007 | -2.381223172 | 0.00000003800447879 |

Table 8: The energy spectra $E_{n,l}$(eV) and expectation values of $\langle r^{-2}\rangle_{nl}$, $\langle V\rangle_{nl}$, $\langle T\rangle_{nl}$ and $\langle p^2\rangle_{nl}$ corresponding to the shifted Deng – Fan molecular potential with various $n$ and $l$ quantum numbers for CuLi diatomic molecule.

| CuLi | | | | | | |
|---|---|---|---|---|---|---|
| $n$ | $l$ | $\langle r^{-2}\rangle_{nl}$ (Å$^{-2}$) | $\langle V\rangle_{nl}$(eV) | $\langle T\rangle_{nl}$(eV) | $E_{n,l}$(eV) | $\langle p^2\rangle_{nl}$((eV/c)$^2$) |
| 0 | 0 | 0.2051863926 | -1.726538140 | 0.01335673701 | -1.713181403 | 0.000000001730641698 |
| 0 | 1 | 0.2051829896 | -1.726537792 | 0.01349341636 | -1.713044375 | 0.000000001748351337 |
| 0 | 2 | 0.2051761839 | -1.726537089 | 0.01376676145 | -1.712770327 | 0.000000001783768850 |
| 0 | 3 | 0.2051659756 | -1.726536017 | 0.01417674501 | -1.712359272 | 0.000000001836890706 |
| 0 | 4 | 0.2051523653 | -1.726534557 | 0.01472332619 | -1.711811231 | 0.000000001907711609 |
| 0 | 5 | 0.2051353534 | -1.726532680 | 0.01540645050 | -1.711126230 | 0.000000001996224500 |
| 1 | 0 | 0.2031152063 | -1.699615547 | 0.03936161311 | -1.660253934 | 0.000000005100111570 |
| 1 | 1 | 0.2031118233 | -1.699614511 | 0.03949622109 | -1.660118290 | 0.000000005117552820 |
| 1 | 2 | 0.2031050576 | -1.699612432 | 0.03976542352 | -1.659847008 | 0.000000005152433567 |
| 1 | 3 | 0.2030949095 | -1.699609296 | 0.04016919337 | -1.659440102 | 0.000000005204750306 |
| 1 | 4 | 0.2030813792 | -1.699605083 | 0.04070749005 | -1.658897593 | 0.000000005274497781 |
| 1 | 5 | 0.2030644674 | -1.699599767 | 0.04138025948 | -1.658219508 | 0.000000005361668984 |
| 2 | 0 | 0.2010541441 | -1.672696317 | 0.06445435167 | -1.608241965 | 0.000000008351395147 |
| 2 | 1 | 0.2010507811 | -1.672694599 | 0.06458690176 | -1.608107698 | 0.000000008368569755 |
| 2 | 2 | 0.2010440551 | -1.672691157 | 0.06485198851 | -1.607839169 | 0.000000008402917229 |
| 2 | 3 | 0.2010339666 | -1.672685977 | 0.06524958508 | -1.607436392 | 0.000000008454434093 |
| 2 | 4 | 0.2010205158 | -1.672679040 | 0.06577965119 | -1.606899388 | 0.000000008523115128 |
| 2 | 5 | 0.2010037034 | -1.672670317 | 0.06644213315 | -1.606228184 | 0.000000008608953376 |
| 3 | 0 | 0.1990031303 | -1.645782651 | 0.08864162297 | -1.557141028 | 0.000000011485356700 |
| 3 | 1 | 0.1989997870 | -1.645780258 | 0.08877212853 | -1.557008130 | 0.000000011502266400 |
| 3 | 2 | 0.1989931006 | -1.645775466 | 0.08903312631 | -1.556742340 | 0.000000011536084060 |
| 3 | 3 | 0.1989830713 | -1.645768262 | 0.08942458968 | -1.556343673 | 0.000000011586806240 |
| 3 | 4 | 0.1989696995 | -1.645758626 | 0.08994647864 | -1.555812147 | 0.000000011654427760 |
| 3 | 5 | 0.1989529858 | -1.645746531 | 0.09059873991 | -1.555147791 | 0.000000011738941710 |

Table 9: The energy spectra $E_{n,l}$(eV) and expectation values of $\langle r^{-2}\rangle_{nl}$, $\langle V\rangle_{nl}$, $\langle T\rangle_{nl}$ and $\langle p^2\rangle_{nl}$ corresponding to the shifted Deng – Fan molecular potential with various $n$ and $l$ quantum numbers for TiC diatomic molecule.

| TiC | | | | | | |
|---|---|---|---|---|---|---|
| $n$ | $l$ | $\langle r^{-2}\rangle_{nl}$ (Å$^{-2}$) | $\langle V\rangle_{nl}$(eV) | $\langle T\rangle_{nl}$(eV) | $E_{n,l}$(eV) | $\langle p^2\rangle_{nl}$((eV/c)$^2$) |
| 0 | 0 | 0.3661250125 | -2.640370623 | 0.01948391604 | -2.620886707 | 0.000000003874270721 |
| 0 | 1 | 0.3661222219 | -2.640370325 | 0.01964294249 | -2.620727383 | 0.000000003905892266 |
| 0 | 2 | 0.3661166409 | -2.640369725 | 0.01996098811 | -2.620408737 | 0.000000003969133908 |
| 0 | 3 | 0.3661082695 | -2.640368815 | 0.02043803834 | -2.619930777 | 0.000000004063992751 |
| 0 | 4 | 0.3660971079 | -2.640367585 | 0.02107407133 | -2.619293514 | 0.000000004190464450 |
| 0 | 5 | 0.3660831562 | -2.640366021 | 0.02186905794 | -2.618496963 | 0.000000004348543214 |
| 1 | 0 | 0.3633939055 | -2.601112815 | 0.05750623948 | -2.543606576 | 0.00000001143480291 |
| 1 | 1 | 0.3633911255 | -2.601111925 | 0.05766348502 | -2.543448440 | 0.00000001146607033 |
| 1 | 2 | 0.3633855656 | -2.601110139 | 0.05797796886 | -2.543132171 | 0.00000001152860373 |
| 1 | 3 | 0.3633772259 | -2.601107453 | 0.05844967650 | -2.542657776 | 0.00000001162240023 |
| 1 | 4 | 0.3633661064 | -2.601103853 | 0.05907858620 | -2.542025267 | 0.00000001174745551 |
| 1 | 5 | 0.3633522075 | -2.601099327 | 0.05986466896 | -2.541234658 | 0.00000001190376379 |
| 2 | 0 | 0.3606711257 | -2.561857832 | 0.09429777009 | -2.467560061 | 0.00000001875059865 |
| 2 | 1 | 0.3606683563 | -2.561856352 | 0.09445324196 | -2.467403110 | 0.00000001878151338 |
| 2 | 2 | 0.3606628174 | -2.561853389 | 0.09476417849 | -2.467089211 | 0.00000001884334142 |
| 2 | 3 | 0.3606545092 | -2.561848936 | 0.09523056525 | -2.466618371 | 0.00000001893607989 |
| 2 | 4 | 0.3606434318 | -2.561842982 | 0.09585238059 | -2.465990601 | 0.000000001905972449 |
| 2 | 5 | 0.3606295854 | -2.561835512 | 0.09662959567 | -2.465205916 | 0.00000001921426948 |
| 3 | 0 | 0.3579566334 | -2.522607533 | 0.12986412700 | -2.392743406 | 0.00000002582277525 |
| 3 | 1 | 0.3579538744 | -2.522605468 | 0.13001783240 | -2.392587636 | 0.00000002585333873 |
| 3 | 2 | 0.3579483564 | -2.522601335 | 0.13032523590 | -2.392276099 | 0.00000002591446426 |
| 3 | 3 | 0.3579400796 | -2.522595126 | 0.13078632340 | -2.391808803 | 0.00000002600614899 |
| 3 | 4 | 0.3579290441 | -2.522586831 | 0.13140107310 | -2.391185758 | 0.00000002612838863 |
| 3 | 5 | 0.3579152500 | -2.522576436 | 0.13216945650 | -2.390406979 | 0.00000002628117748 |

Table 10: The energy spectra $E_{n,l}$(eV) and expectation values of $\langle r^{-2}\rangle_{nl}$, $\langle V\rangle_{nl}$, $\langle T\rangle_{nl}$ and $\langle p^2\rangle_{nl}$ corresponding to the shifted Deng – Fan molecular potential with various $n$ and $l$ quantum numbers for ScF diatomic molecule.

| ScF | | | | | | |
|---|---|---|---|---|---|---|
| $n$ | $l$ | $\langle r^{-2}\rangle_{nl}$ (Å$^{-2}$) | $\langle V\rangle_{nl}$(eV) | $\langle T\rangle_{nl}$(eV) | $E_{n,l}$(eV) | $\langle p^2\rangle_{nl}$((eV/c)$^2$) |
| 0 | 0 | 0.3576312579 | -5.826166085 | 0.02373630190 | -5.802429783 | 0.000000006563759843 |
| 0 | 1 | 0.3576302559 | -5.826165961 | 0.02384808664 | -5.802317874 | 0.000000006594671489 |
| 0 | 2 | 0.3576282519 | -5.826165712 | 0.02407165424 | -5.802094058 | 0.000000006656494263 |
| 0 | 3 | 0.3576252459 | -5.826165336 | 0.02440700094 | -5.801758335 | 0.000000006749227124 |
| 0 | 4 | 0.3576212380 | -5.826164830 | 0.02485412109 | -5.801310709 | 0.000000006872868511 |
| 0 | 5 | 0.3576162280 | -5.826164190 | 0.02541300717 | -5.800751183 | 0.000000007027416345 |
| 1 | 0 | 0.3560483733 | -5.778498652 | 0.07056789853 | -5.707930754 | 0.00000001951402289 |
| 1 | 1 | 0.3560473737 | -5.778498281 | 0.07067894077 | -5.707819340 | 0.00000001954472921 |
| 1 | 2 | 0.3560453743 | -5.778497537 | 0.07090102338 | -5.707596514 | 0.00000001960614134 |
| 1 | 3 | 0.3560423753 | -5.778496419 | 0.07123414260 | -5.707262277 | 0.00000001969825824 |
| 1 | 4 | 0.3560383767 | -5.778494925 | 0.07167829282 | -5.706816632 | 0.00000001982107836 |
| 1 | 5 | 0.3560333784 | -5.778493049 | 0.07223346652 | -5.706259583 | 0.00000001997459962 |
| 2 | 0 | 0.3544684849 | -5.730832405 | 0.1165656282 | -5.614266777 | 0.00000003223369810 |
| 2 | 1 | 0.3544674876 | -5.730831788 | 0.1166759298 | -5.614155858 | 0.00000003226419962 |
| 2 | 2 | 0.3544654929 | -5.730830552 | 0.1168965312 | -5.613934020 | 0.00000003232520214 |
| 2 | 3 | 0.3544625009 | -5.730828695 | 0.1172274286 | -5.613601267 | 0.00000003241670464 |
| 2 | 4 | 0.3544585115 | -5.730826215 | 0.1176686163 | -5.613157599 | 0.00000003253870556 |
| 2 | 5 | 0.3544535249 | -5.730823109 | 0.1182200870 | -5.612603022 | 0.00000003269120282 |
| 3 | 0 | 0.3528915839 | -5.683168130 | 0.1617318560 | -5.521436274 | 0.00000004472343947 |
| 3 | 1 | 0.3528905888 | -5.683167267 | 0.1618414188 | -5.521325848 | 0.00000004475373671 |
| 3 | 2 | 0.3528885988 | -5.683165540 | 0.1620605427 | -5.521104998 | 0.00000004481433066 |
| 3 | 3 | 0.3528856137 | -5.683162948 | 0.1623892238 | -5.520773724 | 0.00000004490522030 |
| 3 | 4 | 0.3528816337 | -5.683159487 | 0.1628274566 | -5.520332030 | 0.00000004502640408 |
| 3 | 5 | 0.3528766587 | -5.683155154 | 0.1633752336 | -5.519779920 | 0.00000004517787994 |

Table 11: The energy spectra $E_{n,l}$(eV) and expectation values of $\langle r^{-2}\rangle_{nl}$, $\langle V\rangle_{nl}$, $\langle T\rangle_{nl}$ and $\langle p^2\rangle_{nl}$ corresponding to the shifted Deng – Fan molecular potential with various $n$ and $l$ quantum numbers for CO diatomic molecule.

| CO | | | | | | |
|---|---|---|---|---|---|---|
| $n$ | $l$ | $\langle r^{-2}\rangle_{nl}$ (Å$^{-2}$) | $\langle V\rangle_{nl}$(eV) | $\langle T\rangle_{nl}$(eV) | $E_{n,l}$(eV) | $\langle p^2\rangle_{nl}$((eV/c)$^2$) |
| 0 | 0 | 0.8986072440 | -11.15293849 | 0.07218859438 | -11.08074990 | 0.00000001025186307 |
| 0 | 1 | 0.8986005888 | -11.15293751 | 0.07273513515 | -11.08020237 | 0.00000001032948006 |
| 0 | 2 | 0.8985872784 | -11.15293553 | 0.07382819238 | -11.07910733 | 0.00000001048471058 |
| 0 | 3 | 0.8985673131 | -11.15293253 | 0.07546771743 | -11.07746481 | 0.00000001071754773 |
| 0 | 4 | 0.8985406932 | -11.15292847 | 0.07765363735 | -11.07527483 | 0.00000001102798114 |
| 0 | 5 | 0.8985074193 | -11.15292331 | 0.08038585487 | -11.07253746 | 0.00000001141599701 |
| 1 | 0 | 0.8921770233 | -11.00761861 | 0.2134564857 | -10.79416212 | 0.00000003031402235 |
| 1 | 1 | 0.8921703933 | -11.00761567 | 0.2139971556 | -10.79361852 | 0.00000003039080559 |
| 1 | 2 | 0.8921571335 | -11.00760979 | 0.2150784712 | -10.79253132 | 0.00000003054436862 |
| 1 | 3 | 0.8921372439 | -11.00760093 | 0.2167003841 | -10.79090055 | 0.000000003074470458 |
| 1 | 4 | 0.8921107251 | -11.00758906 | 0.2188628217 | -10.78872624 | 0.00000003108180315 |
| 1 | 5 | 0.8920775774 | -11.00757414 | 0.2215656871 | -10.78600845 | 0.00000003146565058 |
| 2 | 0 | 0.8857666620 | -10.86230811 | 0.3506879661 | -10.51162014 | 0.00000004980295076 |
| 2 | 1 | 0.8857600572 | -10.86230323 | 0.3512227893 | -10.51108044 | 0.00000004987890367 |
| 2 | 2 | 0.8857468476 | -10.86229347 | 0.3522924116 | -10.51000105 | 0.00000005003080608 |
| 2 | 3 | 0.8857270335 | -10.86227879 | 0.3538967849 | -10.50838200 | 0.00000005025865114 |
| 2 | 4 | 0.8857006153 | -10.86225916 | 0.3560358367 | -10.50622332 | 0.00000005056242859 |
| 2 | 5 | 0.8856675933 | -10.86223453 | 0.3587094709 | -10.50352506 | 0.00000005094212474 |
| 3 | 0 | 0.8793760645 | -10.71701318 | 0.4839017022 | -10.23311147 | 0.000000006872129920 |
| 3 | 1 | 0.8793694847 | -10.71700637 | 0.4844307027 | -10.23257567 | 0.000000006879642520 |
| 3 | 2 | 0.8793563252 | -10.71699275 | 0.4854886797 | -10.23150407 | 0.000000006894667381 |
| 3 | 3 | 0.8793365863 | -10.71697228 | 0.4870755853 | -10.22989669 | 0.000000006917203820 |
| 3 | 4 | 0.8793102682 | -10.71694493 | 0.4891913475 | -10.22775359 | 0.000000006947250817 |
| 3 | 5 | 0.8792773714 | -10.71691066 | 0.4918358705 | -10.22507479 | 0.000000006984807009 |

Table 12: The energy spectra $E_{n,l}$(eV) and expectation values of $\langle r^{-2}\rangle_{nl}$, $\langle V\rangle_{nl}$, $\langle T\rangle_{nl}$ and $\langle p^2\rangle_{nl}$ corresponding to the shifted Deng – Fan molecular potential with various $n$ and $l$ quantum numbers for I$_2$ diatomic molecule.

| I$_2$ | | | | | | |
|---|---|---|---|---|---|---|
| $n$ | $l$ | $\langle r^{-2}\rangle_{nl}$ (Å$^{-2}$) | $\langle V\rangle_{nl}$(eV) | $\langle T\rangle_{nl}$(eV) | $E_{n,l}$(eV) | $\langle p^2\rangle_{nl}$((eV/c)$^2$) |
| 0 | 0 | 0.3141777893 | -1.548880374 | 0.006690598538 | -1.542189775 | 0.000000008787785678 |
| 0 | 1 | 0.3141777763 | -1.548880370 | 0.006711293097 | -1.542169077 | 0.000000008814967005 |
| 0 | 2 | 0.3141777502 | -1.548880363 | 0.006752682209 | -1.542127681 | 0.000000008869329653 |
| 0 | 3 | 0.3141777110 | -1.548880353 | 0.006814765866 | -1.542065587 | 0.000000008950873608 |
| 0 | 4 | 0.3141776588 | -1.548880338 | 0.006897544051 | -1.541982794 | 0.000000009059598851 |
| 0 | 5 | 0.3141775936 | -1.548880321 | 0.007001016743 | -1.541879304 | 0.000000009195505352 |
| 1 | 0 | 0.3139633665 | -1.535441132 | 0.01989603707 | -1.515545095 | 0.00000002613250648 |
| 1 | 1 | 0.3139633534 | -1.535441121 | 0.01991671044 | -1.515524411 | 0.00000002615965998 |
| 1 | 2 | 0.3139633273 | -1.535441100 | 0.01995805718 | -1.515483043 | 0.00000002621396696 |
| 1 | 3 | 0.3139632882 | -1.535441068 | 0.02002007727 | -1.515420991 | 0.00000002629542743 |
| 1 | 4 | 0.3139632360 | -1.535441026 | 0.02010277070 | -1.515338255 | 0.00000002640404136 |
| 1 | 5 | 0.3139631708 | -1.535440973 | 0.02020613746 | -1.515234835 | 0.00000002653980871 |
| 2 | 0 | 0.3137489828 | -1.522001922 | 0.03286772554 | -1.489134196 | 0.00000004317020758 |
| 2 | 1 | 0.3137489697 | -1.522001904 | 0.03288837772 | -1.489113527 | 0.00000004319733326 |
| 2 | 2 | 0.3137489437 | -1.522001869 | 0.03292968209 | -1.489072187 | 0.00000004325158460 |
| 2 | 3 | 0.3137489045 | -1.522001816 | 0.03299163864 | -1.489010177 | 0.00000004333296160 |
| 2 | 4 | 0.3137488524 | -1.522001745 | 0.03307424734 | -1.488927498 | 0.00000004344146423 |
| 2 | 5 | 0.3137487872 | -1.522001657 | 0.03317750818 | -1.488824149 | 0.00000004357709247 |
| 3 | 0 | 0.3135346382 | -1.508562765 | 0.04560572790 | -1.462957037 | 0.00000005990097302 |
| 3 | 1 | 0.3135346252 | -1.508562741 | 0.04562635891 | -1.462936382 | 0.00000005992807088 |
| 3 | 2 | 0.3135345991 | -1.508562691 | 0.04566762093 | -1.462895070 | 0.00000005998226659 |
| 3 | 3 | 0.3135345600 | -1.508562617 | 0.04572951394 | -1.462833103 | 0.00000006006356014 |
| 3 | 4 | 0.3135345078 | -1.508562518 | 0.04581203793 | -1.462750480 | 0.00000006017195151 |
| 3 | 5 | 0.3135344426 | -1.508562395 | 0.04591519287 | -1.462647202 | 0.000000006030744067 |

Table 13: The energy spectra $E_{n,l}$(eV)) and expectation values of $\langle r^{-2}\rangle_{nl}$, $\langle V\rangle_{nl}$, $\langle T\rangle_{nl}$ and $\langle p^2\rangle_{nl}$ corresponding to the shifted Deng – Fan molecular potential with various $n$ and $l$ quantum numbers for HCl diatomic molecule.

| HCl | | | | | | |
|---|---|---|---|---|---|---|
| $n$ | $l$ | $\langle r^{-2}\rangle_{nl}$ (Å$^{-2}$) | $\langle V\rangle_{nl}$(eV) | $\langle T\rangle_{nl}$(eV) | $E_{n,l}$(eV) | $\langle p^2\rangle_{nl}$((eV/c)$^2$) |
| 0 | 0 | 0.6729621729 | -4.51690083 | 0.09985343026 | -4.417047400 | 0.000000002025829696 |
| 0 | 1 | 0.6728782692 | -4.516880401 | 0.1027030746 | -4.414177327 | 0.000000002083643376 |
| 0 | 2 | 0.6727104869 | -4.516838475 | 0.1084002209 | -4.408438254 | 0.000000002199227267 |
| 0 | 3 | 0.6724588767 | -4.516772914 | 0.1169405856 | -4.399832328 | 0.000000002372494468 |
| 0 | 4 | 0.6721235140 | -4.516680515 | 0.1283177476 | -4.388362768 | 0.000000002603314708 |
| 0 | 5 | 0.6717044997 | -4.516557014 | 0.1425231531 | -4.374033861 | 0.000000002891514444 |
| 1 | 0 | 0.6541370829 | -4.312705846 | 0.2844207609 | -4.028285085 | 0.000000005770337805 |
| 1 | 1 | 0.6540545111 | -4.312645806 | 0.2871505060 | -4.025495300 | 0.000000005825718964 |
| 1 | 2 | 0.6538893922 | -4.312524683 | 0.2926078961 | -4.019916787 | 0.000000005936438675 |
| 1 | 3 | 0.6536417757 | -4.312340390 | 0.3007887327 | -4.011551657 | 0.000000006102411758 |
| 1 | 4 | 0.6533117360 | -4.312089800 | 0.3116867218 | -4.000403079 | 0.000000006323510522 |
| 1 | 5 | 0.6528993720 | -4.311768751 | 0.3252934789 | -3.986475272 | 0.000000006599564860 |
| 2 | 0 | 0.6355581099 | -4.108703031 | 0.4496798254 | -3.659023206 | 0.000000009123119173 |
| 2 | 1 | 0.6354768453 | -4.108604414 | 0.4522917545 | -3.656312659 | 0.000000009176110077 |
| 2 | 2 | 0.6353143405 | -4.108406160 | 0.4575135541 | -3.650892606 | 0.000000009282050121 |
| 2 | 3 | 0.6350706442 | -4.108106233 | 0.4653411087 | -3.642765125 | 0.000000009440855807 |
| 2 | 4 | 0.6347458292 | -4.107701581 | 0.4757682487 | -3.631933332 | 0.000000009652401968 |
| 2 | 5 | 0.6343399928 | -4.107188138 | 0.4887867552 | -3.618401383 | 0.000000009916521859 |
| 3 | 0 | 0.6172203112 | -3.905014265 | 0.5960065465 | -3.309007719 | 0.00000001209180053 |
| 3 | 1 | 0.6171403299 | -3.904878073 | 0.5985026908 | -3.306375382 | 0.00000001214244239 |
| 3 | 2 | 0.6169803913 | -3.904604693 | 0.6034929613 | -3.301111732 | 0.00000001224368516 |
| 3 | 3 | 0.6167405430 | -3.904192138 | 0.6109733235 | -3.293218814 | 0.00000001239544700 |
| 3 | 4 | 0.6164208569 | -3.903637427 | 0.6209377297 | -3.282699698 | 0.00000001259760520 |
| 3 | 5 | 0.6160214283 | -3.902936592 | 0.6333781230 | -3.269558469 | 0.00000001284999631 |

Table 14: The energy spectra $E_{n,l}$(eV) and expectation values of $\langle r^{-2}\rangle_{nl}$, $\langle V\rangle_{nl}$, $\langle T\rangle_{nl}$ and $\langle p^2\rangle_{nl}$ corresponding to the shifted Deng – Fan molecular potential with various $n$ and $l$ quantum numbers for LiH diatomic molecule.

| LiH | | | | | | |
|---|---|---|---|---|---|---|
| $n$ | $l$ | $\langle r^{-2}\rangle_{nl}$ (Å$^{-2}$) | $\langle V\rangle_{nl}$(eV) | $\langle T\rangle_{nl}$(eV) | $E_{n,l}$(eV) | $\langle p^2\rangle_{nl}$((eV/c)$^2$) |
| 0 | 0 | 0.4005272888 | -2.463043700 | 0.05111079619 | -2.411932904 | 0.0000000009311574746 |
| 0 | 1 | 0.4004119597 | -2.463025810 | 0.05299498453 | -2.410030826 | 0.0000000009654843915 |
| 0 | 2 | 0.4001813991 | -2.462988399 | 0.05676008694 | -2.406228312 | 0.000000001034078573 |
| 0 | 3 | 0.3998358019 | -2.462928208 | 0.06239956184 | -2.400528646 | 0.000000001136820843 |
| 0 | 4 | 0.3993754605 | -2.462840362 | 0.06990361414 | -2.392936748 | 0.000000001273532750 |
| 0 | 5 | 0.3988007633 | -2.462718382 | 0.07925921597 | -2.383459166 | 0.000000001443976946 |
| 1 | 0 | 0.3859179846 | -2.358655830 | 0.1453935469 | -2.213262283 | 0.000000002648839345 |
| 1 | 1 | 0.3858056701 | -2.358604060 | 0.1471744744 | -2.211429586 | 0.000000002681285014 |
| 1 | 2 | 0.3855811359 | -2.358498954 | 0.1507331617 | -2.207765792 | 0.000000002746118641 |
| 1 | 3 | 0.3852445710 | -2.358337379 | 0.1560632804 | -2.202274099 | 0.000000002843224933 |
| 1 | 4 | 0.3847962590 | -2.358114652 | 0.1631553545 | -2.194959298 | 0.000000002972431251 |
| 1 | 5 | 0.3842365770 | -2.357824546 | 0.1719967802 | -2.185827766 | 0.000000003133507976 |
| 2 | 0 | 0.3716418756 | -2.254440393 | 0.2298573322 | -2.024583060 | 0.000000004187635273 |
| 2 | 1 | 0.3715324824 | -2.254356286 | 0.2315381231 | -2.022818163 | 0.000000004218256611 |
| 2 | 2 | 0.3713137877 | -2.254186565 | 0.2348966399 | -2.019289926 | 0.000000004279443450 |
| 2 | 3 | 0.3709859754 | -2.253928223 | 0.2399267594 | -2.014001464 | 0.000000004371084233 |
| 2 | 4 | 0.3705493203 | -2.253576756 | 0.2466193127 | -2.006957443 | 0.000000004493011917 |
| 2 | 5 | 0.3700041884 | -2.253126181 | 0.2549621047 | -1.998164076 | 0.000000004645004329 |
| 3 | 0 | 0.3576878678 | -2.150503074 | 0.3048344908 | -1.845668583 | 0.000000005553599940 |
| 3 | 1 | 0.3575813060 | -2.150388094 | 0.3064181397 | -1.843969955 | 0.000000005582451506 |
| 3 | 2 | 0.3573682715 | -2.150156688 | 0.3095824716 | -1.840574216 | 0.000000005640100603 |
| 3 | 3 | 0.3570489426 | -2.149805962 | 0.3143215608 | -1.835484401 | 0.000000005726439276 |
| 3 | 4 | 0.3566235863 | -2.149331590 | 0.3206265343 | -1.828705056 | 0.000000005841305874 |
| 3 | 5 | 0.3560925575 | -2.148727822 | 0.3284855908 | -1.820242231 | 0.000000005984485391 |

Table 15: The energy spectra $E_{n,l}$(eV) and expectation values of $\langle r^{-2}\rangle_{nl}$, $\langle V\rangle_{nl}$, $\langle T\rangle_{nl}$ and $\langle p^2\rangle_{nl}$ corresponding to the shifted Deng – Fan molecular potential with various $n$ and $l$ quantum numbers for H$_2$ diatomic molecule.

| H$_2$ | | | | | | |
|---|---|---|---|---|---|---|
| $n$ | $l$ | $\langle r^{-2}\rangle_{nl}$ (Å$^{-2}$) | $\langle V\rangle_{nl}$(eV) | $\langle T\rangle_{nl}$(eV) | $E_{n,l}$(eV) | $\langle p^2\rangle_{nl}$((eV/c)$^2$) |
| 0 | 0 | 1.774568810 | -4.566076874 | 0.1714570951 | -4.394619779 | 0.000000001788446691 |
| 0 | 1 | 1.771531580 | -4.565761457 | 0.1858502220 | -4.379911235 | 0.000000001938579529 |
| 0 | 2 | 1.765474347 | -4.565056645 | 0.2144870309 | -4.350569614 | 0.000000002237286362 |
| 0 | 3 | 1.756431349 | -4.563815891 | 0.2570707506 | -4.306745141 | 0.000000002681471613 |
| 0 | 4 | 1.744453407 | -4.561822853 | 0.3131614720 | -4.248661381 | 0.000000003266546643 |
| 0 | 5 | 1.729607288 | -4.558795463 | 0.3821823065 | -4.176613157 | 0.000000003986494004 |
| 1 | 0 | 1.634352381 | -4.210060596 | 0.4622376494 | -3.747822947 | 0.000000004821540888 |
| 1 | 1 | 1.631513452 | -4.209200581 | 0.4749238187 | -3.734276762 | 0.000000004953868672 |
| 1 | 2 | 1.625851538 | -4.207413746 | 0.5001588129 | -3.707254933 | 0.000000005217091619 |
| 1 | 3 | 1.617398329 | -4.204567765 | 0.5376698791 | -3.666897886 | 0.000000005608364680 |
| 1 | 4 | 1.606200866 | -4.200467301 | 0.5870526831 | -3.613414618 | 0.000000006123470294 |
| 1 | 5 | 1.592320965 | -4.194857709 | 0.6477769379 | -3.547080771 | 0.000000006756877110 |
| 2 | 0 | 1.502413174 | -3.856915237 | 0.6925636029 | -3.164351634 | 0.000000007224040995 |
| 2 | 1 | 1.499757136 | -3.855574730 | 0.7036755227 | -3.151899207 | 0.000000007339947988 |
| 2 | 2 | 1.494459835 | -3.852833353 | 0.7257730003 | -3.127060352 | 0.000000007570443907 |
| 2 | 3 | 1.486550642 | -3.848571538 | 0.7586050835 | -3.089966454 | 0.000000007912911103 |
| 2 | 4 | 1.476073157 | -3.842612807 | 0.8017997322 | -3.040813074 | 0.000000008363468873 |
| 2 | 5 | 1.463084681 | -3.834727128 | 0.8548689712 | -2.979858157 | 0.000000008917027213 |
| 3 | 0 | 1.378068693 | -3.508172439 | 0.8676585950 | -2.640513844 | 0.000000009050434118 |
| 3 | 1 | 1.375581606 | -3.506407648 | 0.8773154173 | -2.629092231 | 0.000000009151163177 |
| 3 | 2 | 1.370621142 | -3.502823493 | 0.8965126876 | -2.606310805 | 0.000000009351407410 |
| 3 | 3 | 1.363214554 | -3.497311883 | 0.9250192810 | -2.572292602 | 0.000000009648755984 |
| 3 | 4 | 1.353402305 | -3.489713299 | 0.9624925328 | -2.527220766 | 0.000000010039634610 |
| 3 | 5 | 1.341237578 | -3.479819850 | 1.008482962 | -2.471336889 | 0.000000010519354800 |

## 4. Conclusion

In this work, we have studied the approximate analytical solutions of the Schrödinger equation for the shifted Deng – Fan molecular potential with a suitable approximation to the centrifugal term using the proper quantization rules. The explicit bound state energy equation for this potential are obtained. Furthermore, we applied Hellman-Feynman theorem on the solutions and obtained the formula for the expectation values of $r^{-2}$, $p^2$, $T$ and $V$. The solutions obtained have been used to calculate the numerical values of the energy spectra $E_{n,l}$(eV) and the expectation values of $r^{-2}$, $p^2$, $T$ and $V$ for all the selected diatomic molecules (HCl, LiH, $H_2$, CO, ScH, TiH, VH, CrH, CuLi, TiC, NiC, ScN, ScF & $I_2$) for arbitrary values of n and $l$ quantum numbers.

The results of the energy spectra $E_{n,l}$(eV) and the expectation values are displayed in Tables 2-15. It is found from the tables presented that energies increase as the quantum numbers (n, $l$) of the state increase for all the selected diatomic molecules. These results are in excellent agreement with the available ones in the literature (see results in Ref. [16]). The expectation values of $r^{-2}$ decreases as the quantum numbers (n, $l$) of the state increase while that of $V$, $T$ and $p^2$ increase for all the selected diatomic molecules.

The results obtained are sufficiently accurate and very consistent. The method employed is simple, unique and it is as suitable as other methods for finding bound state energies only. Our results are very useful in mathematical physics, chemical physics, quantum physics, quantum chemistry, high energy physics and other related areas.

**Acknowledgement:** The author(s) thank professors S.H. Dong, Z.Q Ma, W. C. Qiang, F.A. Serrano and M. C. Irisson for making their papers available which have greatly improve the quality of this work.

**Conflict of Interest:** I declare that there is no conflict of interest regarding this manuscript

**References**

[1] A.A. Zevitsas, J. Am. Chem. Soc., **113** (1991) 4755.

[2] P.G. Hajigeorgiou, J. Mol. Spec., **263** (2010) 101.

[3] G. Herzberg, *Molecular spectra and structure I: Spectra of Diatomic molecules,* New York: Van Nostrand (1950).

[4] P. A. Fraser and W. R. Jarmain, Proc. Phys. Soc. **66**. (1953) 114.

[5] Y. P .Varshni, Rev. Mod. Phys. **31**(1959) 839

[6] J. F. Harrison, Chem. Rev. **100**, (2000) 679.

[7] Z. Rong, H. G. Kjaergaard and M. L. Sage, Mol. Phys. **101**(2003)2285.

[8] A .T. Royappa, V. Suri and J. R. McDonough, J. Mol. Struc. **787** (2006) 209.

[9] S .H. Dong, *Factorization Method in Quantum Mechanics, Fundamental Theories in Physics*, 150, Netherlands, Springer (2007).

[10] Y. P. Varshni Chem. Phys. **353** (2008)32.

[11] K. J. Oyewumi and K.D. Sen, J. Math. **50** (2012) 1039

[12] K. J. Oyewumi, O. J. Oluwadare, K.D. Sen and O. A. Babalola, J. Math. Chem. **51**(2013) 976.

[13] S. M. Ikhdair and B.J. Falaye, Chem. Phys. **421**(2013) 84.

[14] C.A. Onate, Chin. J. Phys. **54** (2016) 165.

[15] M. Hamzavi, S.M Ikhdair and K.E Thylwe, J. Math. Chem. **51** (2013) 227.

[16] K. J. Oyewumi, B.J. Falaye, C.A. Onate, O. J. Oluwadare and W.A. Yahya, Mol. Phys. **112** (2014) 127.

[17] Z.Q Ma and B.W. Xu, Int. J. Mod. Phys. E, **14** (2005a) 599

[18] Z.Q Ma and B.W. Xu, Euro. Phys. Lett. **69** (2005 b) 685

[19] C.N. Yang, *Monopoles in Quantum Field Theory*. Proc. of the Monopole Meeting (Trieste, Italy), ed. N.S Craigie, P Goddard and W. Nahm, Singapore: World Scientific (1982) 237.

[20] W. C. Qiang and S.H. Dong, Eur. Phys. Lett., **89** (2010) 10003.

[21] W. C. Qiang and S.H. Dong, Phys. Lett. A. **363** (2007) 169.

[22] S.H. Dong, D. Morales and J. Garcia-Ravelo, Int. J. Mod. Phys. E **16** (2007) 189.

[23] X.Y. Gu, S.H. Dong and Z.Q Ma, J. Phys. A: Math Theor. **42** (2009) 035303.

[24] X.Y. Gu and S.H. Dong, J. Math. Chem. **49** (2011) 2053.

[25] S.H. Dong and M. C. Irisson, J. Math. Chem. **50** (2012) 881.

[26] F.A. Serrano, X.Y Gu, and S.H. Dong, J. Math. Phys. **51**(2010) 082103.

[27] F.A. Serrano, M. C. Irisson and S.H. Dong, Ann. Phys.**523** (2011) 771.

[28] R. L. Greene and C. Aldrich, Phys. Rev. A. **143** (1976) 2363.

[29] S.H. Dong, and X.Y Gu, J. Phys. Conf. Series **96** (2008) 012109.

[30] G. F. Wei, C.Y. Long and S.H. Dong, Phys. Lett. A, **372** (2008a) 2592.

[31] G. F. Wei, C.Y. Long, Duan X Y and S .H. Dong Phys. Scr., **78** (2008b) 035001.

[32] G. Hellmann, Einführung in die Quantenchemie, Denticke, Vienna, (1937).

[33] R .P. Feynman, Phys. Rev. **56** (1939) 340.

[34] G. Marc and W. G. McMillan, Adv. Chem. Phys. **58** (1985) 209.

[35] D. Popov, Int. J. Quant. Chem., **69** (1998) 159.

[36] D. Popov, Czech. J. Phys. **49** (1999) 1121.

[37] D. Popov, J. Phys. A. Math. Gen. 34 (2001) 5283.

[38] M. E Grypeos, C. G. Koutroulos, K. J. Oyewumi and Th. A. Petridou, J. Phys. A: Math. Gen. **37** (2004) 7895.

[39] K. J. Oyewumi, Th. A. Petridou, M. E Grypeos and C. G. Koutroulos, *Proceedings of the 3rd International Workshop on Contemporary Problems in Mathematical Physics,* COPROMAPH 3 held on the 1st - 7th Nov. 2003: eds. Profs. J. Govaerts, M. N. Hounkonnu and A. Z. Msezane, (2004) 336

[40] K.J. Oyewumi, Found. Phys. Lett.18 (2005), 75, K. J 2008 Proceedings of the fifth International Workshop on Contemporary Problems in Mathematical Physics COPROMAPH 5 held on the 27th Oct.− 2nd Nov. 2007 under the auspices of the International Chair in Mathematical Physics and Applications ICMPA, University of Abomey-Calavi, Cotonou, Benin Republic. eds. Profs. J. Govaerts and M. N. Hounkonnu, 193.

[41] D. B. Wallace, *An Introduction to Hellmann-Feynman Theory* M. Sc. Thesis, University of Central Florida, USA (2005).